\documentclass[conference]{IEEEtran}
\IEEEoverridecommandlockouts

\usepackage{cite}
\usepackage{amsmath,amssymb,amsfonts}
\usepackage{graphicx}
\usepackage{textcomp}
\usepackage{xcolor}
\usepackage{url}
\usepackage[hidelinks]{hyperref}

\graphicspath{{}}
\newif\ifshowfixme
\showfixmefalse

\begin{document}

\title{EMA-Based Subspace Tracking for Adaptive\\
Artifact Subspace Reconstruction}

\author{
\IEEEauthorblockN{Vishnu KN}
\IEEEauthorblockA{\textit{Neural Engineering Lab} \\
\textit{Dept. of Biosciences and Bioengineering} \\
\textit{Indian Institute of Technology}\\
Guwahati, India \\
kvishnu@iitg.ac.in}
\and
\IEEEauthorblockN{Cota Navin Gupta}
\IEEEauthorblockA{\textit{Neural Engineering Lab} \\
\textit{Dept. of Biosciences and Bioengineering} \\
\textit{Indian Institute of Technology}\\
Guwahati, India \\
cngupta@iitg.ac.in}
}

\maketitle
\begin{abstract}
Electroencephalogram (EEG) artifact removal remains a critical challenge for real-world brain--computer interface deployment due to non-stationary signal statistics and limited calibration availability. Artifact Subspace Reconstruction (ASR) provides an automated framework for variance-based artifact suppression but relies on a static calibration-derived covariance model, limiting stability under intra-subject drift and increasing sensitivity to parameter mis-specification. This work proposes an adaptive ASR framework
based on exponential moving average (EMA) subspace tracking. By integrating short-horizon covariance estimation with slow recursive assimilation, the method establishes dual adaptation timescales that enable responsiveness to evolving EEG structure while preserving stability against transient artifacts. This formulation further promotes smoother operating characteristics across rejection thresholds, reducing sensitivity to hyperparameter selection in
continuous deployment settings. Evaluated on 24-channel EEG recordings from 10 subjects with prominent blink artifacts across cognitive states, EMA-ASR achieved substantially stronger artifact attenuation than Original and memory-limited ASR variants (53.1\% vs.\ 24.0\% and 26.7\% blink reduction), at the cost of increased reconstruction and spectral deviation, whereas the memory-limited variant performed comparably to Original ASR. These findings position EMA-based adaptive tracking as a lightweight, deployment-oriented approach to continuous EEG artifact removal under non-stationary conditions, with an explicit trade-off between suppression strength and signal preservation.
\end{abstract}

\begin{IEEEkeywords}
EEG, Artifact Subspace Reconstruction, Adaptive Subspace Tracking, Exponential
Moving Average, Non-Stationary EEG, Covariance Adaptation, Real-Time EEG
Processing
\end{IEEEkeywords}

\section{Introduction}

Electroencephalogram (EEG)-based brain--computer interfaces (BCIs) are
increasingly deployed for real-world neural monitoring across professional \cite{pan_fatigue_2024,balam_systematic_2024} and consumer domains, including well-being \cite{riaz_wearable_2024} and gaming applications \cite{keutayeva_neurotechnology_2025}. Key advantages of EEG-BCI systems include portability, high temporal resolution, and affordability, enabling neuromonitoring in natural environments \cite{hazarika_smartphone-based_2022}. However, EEG signals exhibit low signal-to-noise ratio and are highly susceptible to artifacts such as blinks, eye movements, and muscle activity.
These high-amplitude transients obscure neural activity and must be removed for reliable downstream analysis \cite{uriguen_eeg_2015}.

Artifact Subspace Reconstruction (ASR) \cite{kothe_artifact_2015-1} provides an automated, online-capable artifact removal framework based on principal component analysis (PCA). ASR estimates a subject-specific clean covariance model from calibration data and evaluates incoming EEG windows relative to this baseline \cite{chang_evaluation_2020}. Statistically deviant activity
manifests as high-variance components in the principal subspace and is
attenuated through thresholding and signal reconstruction. However, EEG
statistics are inherently non-stationary, raising concerns regarding the long-term stability of a calibration-derived covariance reference. This formulation implicitly assumes statistical stationarity despite multi-timescale variability across neural and artifact components in the signal. Nevertheless, due to its superiority in automation, signal fidelity, and real-time capability, ASR remains widely adopted \cite{gorjan_removal_2022} and is considered a strong candidate for standardized EEG preprocessing pipelines \cite{miyakoshi_artifact_2023}. Despite its popularity, classical ASR has notable limitations. It relies on multi-channel data and a static
clean-covariance model derived from calibration recordings, typically acquired within the same session, necessitating repeated user-specific calibration, posing challenges for consumer-grade deployment.

Several extensions have sought to address these constraints. Embedded ASR (E-ASR) \cite{hazarika_dynamical_2024} relaxes the multi-channel requirement through dynamical embedding of single-channel signals, while Hardware-Oriented Memory-Limited ASR (HMO-ASR) \cite{van_hardware-oriented_2021} introduces iterative covariance updates and computational optimizations for real-time implementation. These approaches primarily target single-channel feasibility and online efficiency, respectively. However, such variants remain
calibration-dependent. Hebbian-ASR \cite{tsai_development_2024} introduces adaptive mixing and threshold learning across subjects, but its segment-wise optimization and computational overhead limit real-time applicability, emphasizing cross-subject accuracy over deployment efficiency.

In this work, we propose an adaptive framework for clean-EEG subspace
tracking, based on exponential moving average (EMA) updates of the clean covariance model. Unlike Hebbian or moment-based iterative schemes that explicitly optimize basis vectors, EMA assimilates statistical evidence through a single adaptation-rate parameter, enabling stable, low-complexity tracking of intra-subject non-stationarity. Signal characteristics are monitored using a rolling 2-s buffer window, introducing dual adaptation timescales: a fast buffer-driven timescale capturing rapid covariance fluctuations, and a slower EMA-governed timescale enforcing gradual, memory-weighted subspace evolution and artifact thresholding. Together, these mechanisms operationalize a lightweight adaptive subspace-tracking framework
that preserves the statistical integrity of the clean model while enabling continuous post-calibration adaptation under non-stationary EEG conditions.

\section{Methods}

\subsection{The Original ASR Algorithm}

ASR operates in two sequential phases: \emph{calibration} and
\emph{processing}. During calibration, a clean covariance model is estimated from artifact-free EEG data. This model defines variance-based amplitude thresholds that characterize the statistical structure of clean neural activity. In the processing phase, incoming EEG is segmented into sliding windows and compared against the calibration model. Each window is projected into the principal component (PC) subspace of the clean covariance. Components whose variance exceeds predefined thresholds are identified as artifactual and pruned. The remaining signal is reconstructed using the clean mixing matrix
\cite{miyakoshi_artifact_2023}.

\subsection{The Adaptive-ASR Algorithm}

Adaptive-ASR extends the Original ASR framework through two complementary update mechanisms:
\begin{enumerate}
\item A short-horizon clean buffer (2\,s) containing windows classified as artifact-free.
\item A slow exponential moving average (EMA) update governing both the
reconstruction matrix and rejection thresholds.
\end{enumerate}

This dual-timescale design enables responsiveness to persistent non-stationary changes while resisting corruption from transient artifacts, thereby preserving ASR's real-time efficiency without incurring the computational burden associated with Hebbian-ASR.

\subsubsection{Calibration Phase}
The calibration phase follows the standard ASR procedure. A Yule--Walker pre-emphasis filter is applied prior to covariance estimation. Clean covariance is robustly estimated using blockwise geometric median smoothing to obtain the baseline model $\Sigma_0$. The mixing matrix is computed as the symmetric matrix square root: $M_0 = \sqrt{\Sigma_0}$. Principal-component rejection thresholds are defined using robust RMS statistics: $T = \mu + k \cdot \sigma$, where $\mu$ and $\sigma$ are robust RMS statistics and $k$ controls rejection aggressiveness (see Fig.~\ref{fig:cutoff}). This establishes the initialization model for subsequent adaptive updates.

\begin{figure*}[t]
\centering
\includegraphics[width=0.9\textwidth]{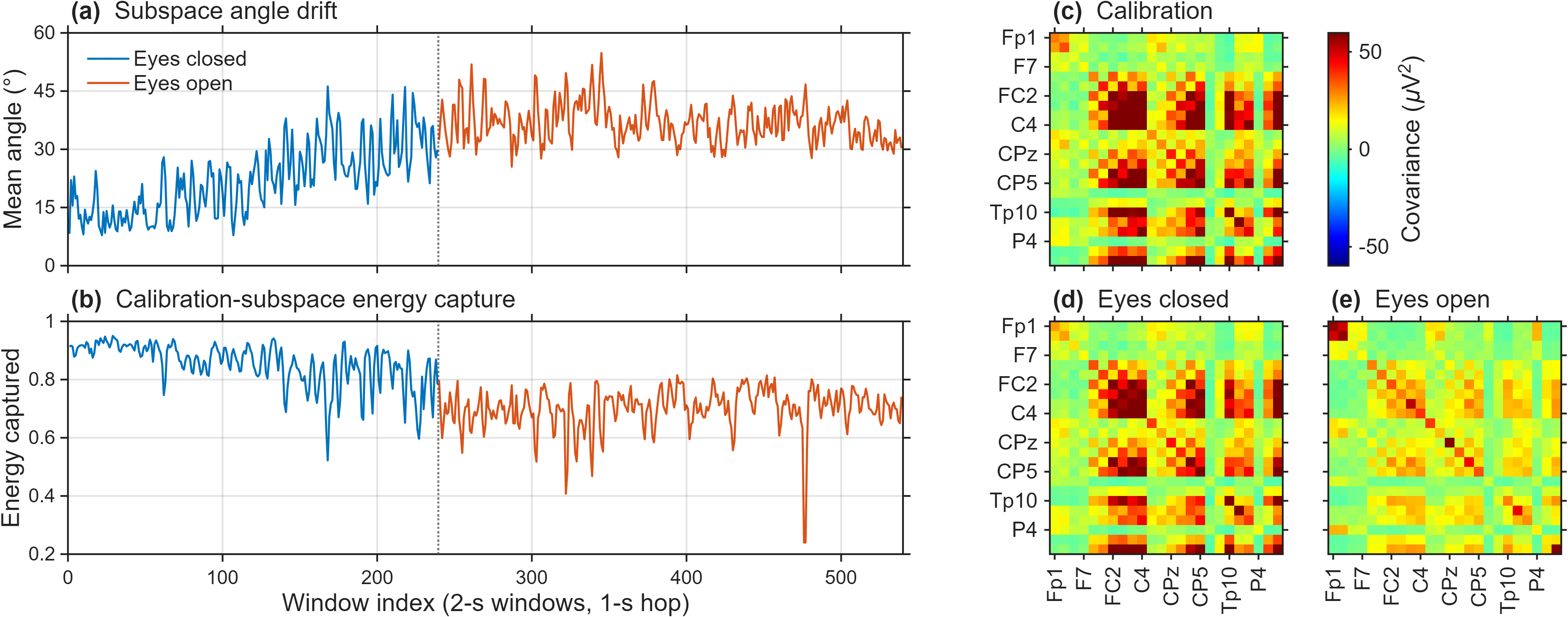}
\caption{Covariance structure, subspace angle drift, and calibration energy capture across cognitive states demonstrate systematic intra-subject non-stationarity and progressive divergence from the calibration-defined clean subspace.}
\label{fig:nonstationarity}
\end{figure*}

\subsubsection{Processing Phase}
During online operation, EEG windows are projected into the PC basis derived from the current mixing matrix. Component amplitudes are compared against adaptive thresholds. Components exceeding thresholds are truncated, and the signal is reconstructed via pseudoinverse projection. To ensure causal real-time operation, a small look-ahead buffer is employed together with raised-cosine blending across window boundaries.

\paragraph{Clean Model Update Mechanism}
Windows not marked for rejection contribute to model adaptation. Accepted windows populate the short-horizon clean buffer. From this buffer, a candidate clean covariance is estimated using generalized inner-product covariance estimation with geometric median smoothing: $\Sigma_{\text{candidate}}$. The corresponding candidate mixing matrix is obtained as: $M_{\text{candidate}} = \sqrt{\Sigma_{\text{candidate}}}$. This candidate basis is assimilated into the adaptive mixing matrix via exponential moving average:

\begin{equation}
M_{t+1} = (1-\alpha) M_t + \alpha M_{\text{candidate}}, \qquad 0 < \alpha < 1
\label{eq:ema_m}
\end{equation}

Similarly, candidate per-component thresholds are computed from the clean buffer: $T_{\text{candidate}} = \mu_{\text{buffer}} + k \cdot
\sigma_{\text{buffer}}$. These are blended with the current adaptive
thresholds as:

\begin{equation}
T_{t+1} = (1-\alpha) T_t + \alpha T_{\text{candidate}}, \qquad 0 < \alpha < 1
\label{eq:ema_t}
\end{equation}

where $\alpha$ controls the adaptation rate. For the present experiments, $\alpha = 0.05$, ensuring gradually decaying statistical memory while maintaining stability against transient fluctuations.

\subsubsection{Dual-Timescale Adaptation Dynamics}
The interaction between the clean buffer and EMA establishes two adaptation timescales. The fast buffer-driven timescale enables rapid assimilation of persistent covariance changes, supporting responsiveness to intra-subject non-stationarity. The slower EMA timescale governs gradual statistical integration, preserving stability and limiting susceptibility to transient artifacts.

\subsection{EEG Data}

The algorithms were evaluated on an EEG dataset containing prominent eye-blink artifacts, collected in prior work 
\cite{hazarika_dynamical_2024,hazarika_-device_2026}. Recordings were obtained from 10 healthy adults (7 male, 3 female; mean age $25.9 \pm 3.36$ years) during a 10-minute resting-state paradigm comprising 5-minute eyes-open and eyes-closed segments. EEG was acquired from 24 scalp electrodes (EasyCap) using an mBrainTrain Smarting amplifier at 500\,Hz sampling rate, with synchronized facial video recorded via the CameraEEG Android application \cite{hazarika_cameraeeg_2023}. Data collection was approved by the IIT Guwahati Human Ethics Committee. Preprocessing included zero-mean normalization, 0.5--40\,Hz band-pass filtering, and 50\,Hz notch filtering. Algorithms were implemented in MATLAB (R2025b) using EEGLAB (v2025.1.0) and evaluated in simulated online mode.

The eyes-closed condition exhibited pronounced alpha enhancement relative to eyes-open recordings, inducing systematic shifts in clean covariance structure. These cognitive-state transitions form the primary intra-subject non-stationarity axis evaluated in this study.

\subsection{Experimental Design and Evaluation}

Three artifact-removal methods were compared:

\begin{itemize}
\item Baseline ASR with static calibration -- ``ASR''
\item HMO-ASR with iterative statistical updates -- ``HMO''
\item Proposed Adaptive ASR with EMA-based updates -- ``EMA''
\end{itemize}

All methods were initialized using identical subject-specific calibration data and matched ASR parameters, including a rejection cut-off of $k = 15$ unless stated otherwise. Calibration segments were drawn from relatively clean eyes-closed data (first 2--42\,s) and used to initialize each algorithm. Performance
metrics were computed per channel and averaged across the 24 scalp locations, followed by aggregation across subjects to obtain cohort-level performance estimates. Pairwise differences between methods were assessed across subjects ($n = 10$) using two-sided exact Wilcoxon signed-rank tests ($p < 0.05$).

\subsection{Ground Truth of EEG Artifacts}

Blink events were manually annotated using synchronized facial video.
Automated detections were additionally performed using amplitude peaks
exceeding six times the mean absolute signal amplitude, with a minimum
inter-blink interval of 250\,ms. Automated counts closely matched video annotations ($\pm 1$ blink discrepancy) \cite{hazarika_dynamical_2024}.

Drift correction via zero-mean centering was applied prior to evaluation. Minimal calibration data (40\,s) were used solely for initialization, constituting a low-calibration regime relative to conventional ASR requirements ($\sim$2\,min of clean data), thereby enabling evaluation of adaptive stability under limited calibration availability.

\subsection{Performance Metrics}

Artifact-removal performance was quantified using blink reduction, defined as the percentage decrease in detected blink events after cleaning. Additional measures included proportion of modified samples and variance attenuation as functions of cut-off parameter $k$. Spectral and statistical integrity were assessed using normalized band-power deviations, Pearson correlation with the raw signal, and relative root mean square error (RRMSE). Together, these
metrics capture both artifact suppression efficacy and preservation of
neurophysiological signal structure.

\section{Results}

\subsection{Intra-Subject Non-stationarity}

\begin{table}[!t]
\centering
\caption{Primary cleaning performance metrics (mean $\pm$ std, $n=10$). $^{*}p<0.05$ vs.\ ASR}
\label{tab:primary}
\begin{tabular}{|c|c|c|c|}
\hline
\textbf{Algorithm} & \textbf{RRMSE} & \textbf{Correlation} & \textbf{Reduction (\%)} \\
\hline
EMA & $0.429 \pm 0.152^{*}$ & $0.861 \pm 0.101^{*}$ & $53.07 \pm 17.83^{*}$ \\
HMO & $0.280 \pm 0.142$ & $0.933 \pm 0.057$ & $26.68 \pm 18.40$ \\
ASR & $0.224 \pm 0.165$ & $0.939 \pm 0.059$ & $24.04 \pm 20.78$ \\
\hline
\end{tabular}
\end{table}

Figure~\ref{fig:nonstationarity} illustrates intra-subject non-stationarity across calibration, eyes-closed, and eyes-open conditions. Covariance heatmaps reveal systematic reorganization of channel correlations across cognitive states, with eyes-open recordings exhibiting attenuated long-range structure relative to calibration and eyes-closed segments. Subspace angle analysis
quantifies this drift, with principal subspace deviations exceeding
30--40\textdegree{} during state transitions, indicating substantial geometric divergence from the calibration basis. Correspondingly, calibration subspace energy capture decreases under non-stationary conditions, reflecting reduced representational adequacy of the static clean model.

Collectively, these results demonstrate that clean EEG covariance structure evolves systematically across cognitive states, leading to measurable subspace misalignment and representational decay when calibration-defined references are assumed fixed.

\subsection{Artifact Suppression Performance}

Table~\ref{tab:primary} summarizes primary cleaning performance using RRMSE, correlation, and blink reduction. EMA-ASR achieved the strongest artifact attenuation, roughly doubling blink reduction relative to Original ASR ($53.07\%$ vs.\ $24.04\%$), but at the cost of higher reconstruction distortion, with increased RRMSE and reduced correlation with the raw signal. Original ASR yielded the highest reconstruction fidelity, reflected by the lowest RRMSE and
highest signal correlation. All three differences between EMA-ASR and Original ASR were statistically significant ($p < 0.05$), as were those between EMA-ASR and HMO-ASR. HMO-ASR did not differ significantly from original ASR on any metric at this operating point.

\begin{table*}[!t]
\centering
\caption{Band power differences after cleaning (mean $\pm$ std, $n=10$). $^{*}p<0.05$ vs.\ ASR}
\label{tab:bandpower}
\begin{tabular}{|c|c|c|c|c|c|}
\hline
\textbf{Algorithm} & $\Delta\delta$ & $\Delta\theta$ & $\Delta\alpha$ & $\Delta\beta$ & $\Delta\gamma$ \\
\hline
EMA & $0.01816 \pm 0.01777^{*}$ & $0.00566 \pm 0.00740^{*}$ & $0.00633 \pm 0.01376^{*}$ & $0.00424 \pm 0.01052^{*}$ & $0.00043 \pm 0.00072^{*}$ \\
HMO & $0.01311 \pm 0.01500$ & $0.00289 \pm 0.00283$ & $0.00311 \pm 0.00501$ & $0.00063 \pm 0.00100$ & $0.00011 \pm 0.00017$ \\
ASR & $0.01338 \pm 0.01640$ & $0.00230 \pm 0.00276$ & $0.00041 \pm 0.00041$ & $0.00031 \pm 0.00032$ & $0.00010 \pm 0.00017$ \\
\hline
\end{tabular}
\end{table*}

\begin{figure*}[t]
\centering
\includegraphics[width=\textwidth]{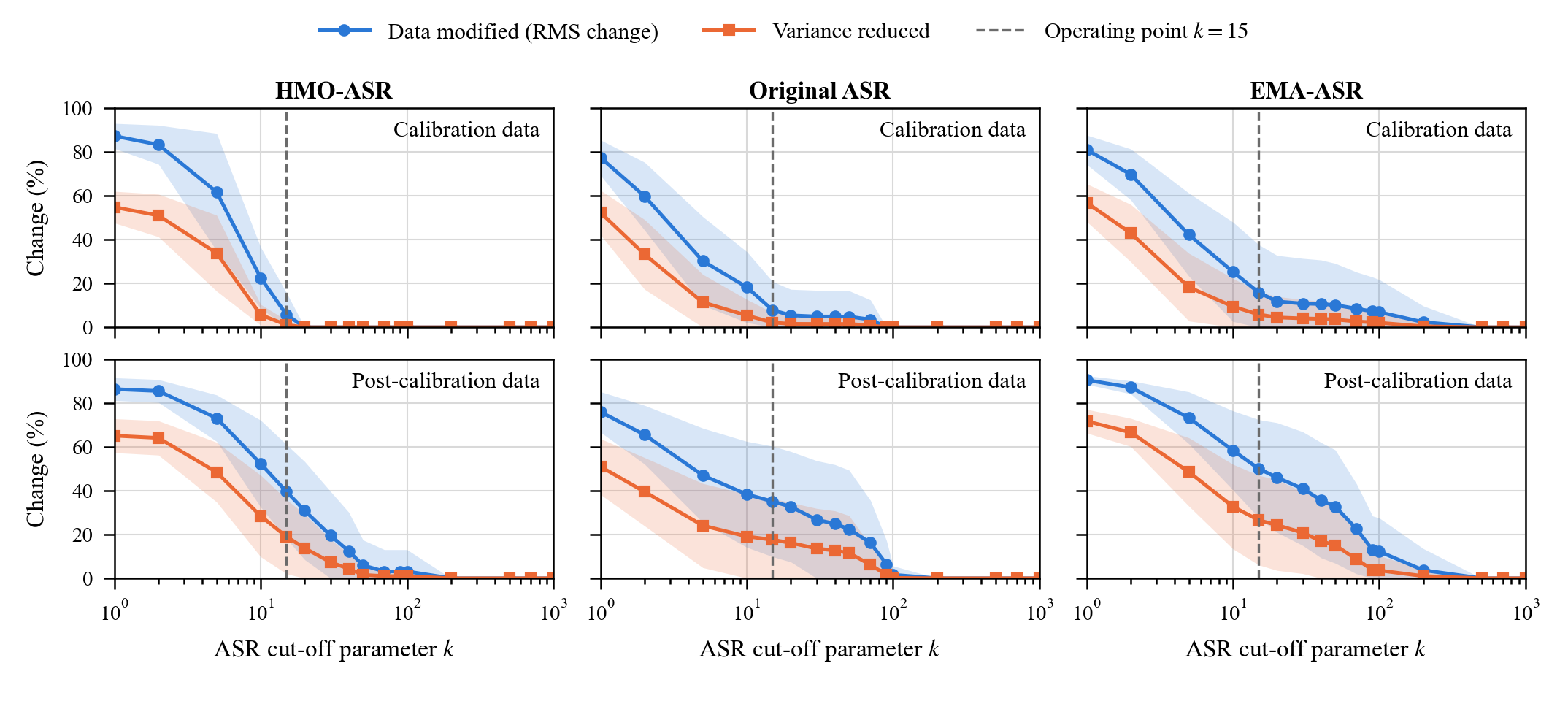}
\caption{Percentage of modified data samples (blue) and variance reduction (red) are plotted as functions of the ASR cutoff parameter $k$ for Original ASR, HMO-ASR, and adaptive (EMA-based) ASR. Results are shown for calibration segments (top row) and full recordings (bottom row), with shaded regions indicating $\pm 1$ standard deviation across subjects. The operating point at $k = 15$ is marked by the vertical dashed line.}
\label{fig:cutoff}
\end{figure*}

\subsection{Spectral Preservation Analysis}

To assess neural signal integrity, band-limited power deviations were computed across canonical EEG bands (Table~\ref{tab:bandpower}). For all methods, the largest deviations occurred in the delta band, consistent with the low-frequency spectral content of blink artifacts. Original ASR produced the smallest perturbations in the theta, alpha, beta, and gamma bands. EMA-ASR
introduced the largest deviations in every band, differing significantly from Original ASR across all five bands ($p < 0.05$), with the most pronounced relative increases in the alpha and beta bands, together with the highest inter-subject variability. HMO-ASR produced deviations between those of original ASR and EMA-ASR in the theta–gamma bands and the smallest delta deviation.

These findings reinforce temporal-domain observations: stronger artifact suppression is accompanied by increased spectral distortion, whereas conservative rejection better preserves intrinsic oscillatory structure.

\subsection{Dependency on Threshold Cut-off Parameter \texorpdfstring{$k$}{k}}

Figure~\ref{fig:cutoff} characterizes algorithm sensitivity to the variance cut-off parameter $k$ using variance reduction and proportion of modified samples. HMO-ASR exhibits a narrow operating regime with steep response transitions, indicating high parameter sensitivity. Original ASR demonstrates a broader but more permissive response profile, while EMA-ASR shows smoother transitions between aggressive and conservative rejection. At the standardized operating point ($k = 15$), all methods modified approximately 25--30\% of data while achieving 15--20\% variance reduction.

\section{Discussion}

\subsection{Adaptive Necessity Under Non-Stationarity}

Covariance and subspace analyses (Fig.~\ref{fig:nonstationarity}) demonstrate that calibration-derived clean models undergo systematic geometric drift across cognitive states, resulting in progressive subspace misalignment and reduced representational energy. In parallel, cutoff sensitivity analysis (Fig.~\ref{fig:cutoff}) indicates that artifact rejection behavior remains strongly dependent on calibration-anchored variance thresholds.

Together, these observations suggest that both the statistical reference (clean covariance) and its rejection boundary ($k$) are susceptible to non-stationary drift. Static calibration therefore constrains ASR within a temporally brittle operating regime, wherein fixed subspaces and thresholds may no longer remain representative of ongoing EEG dynamics.

These findings motivate adaptive subspace tracking mechanisms capable of continuously assimilating evolving covariance evidence while preserving stability against transient perturbations.

\subsection{Suppression--Preservation Trade-off}

Comparative results reveal a consistent suppression--distortion trade-off across algorithms. EMA-ASR maximizes artifact attenuation but introduces greater reconstruction deviation, whereas Original ASR preserves signal structure with minimal distortion but achieves weaker artifact suppression. HMO-ASR occupies an intermediate position that is statistically indistinguishable from original ASR at k = 15; given its steep dependence on k (Fig. \ref{fig:cutoff}), this similarity may not generalize to other cut-offs.

This trade-off reflects fundamental constraints in variance-based artifact rejection: aggressive suppression improves artifact removal but risks attenuating neural activity embedded within high-amplitude transients, while conservative rejection preserves neurophysiological structure at the expense of residual contamination.

Spectral analysis reinforces this dynamic. Original ASR exhibits minimal band-power deviations, particularly in the alpha and beta rhythms critical for cognitive-state interpretation. EMA-ASR introduces significantly larger deviations across all bands, including alpha and beta, indicating that its stronger suppression extends beyond blink-dominated low frequencies. Because RRMSE, correlation, and band-power deviations are computed relative to the uncleaned signal, these measures cannot distinguish removal of residual
artifact from attenuation of neural activity; resolving this requires a
clean-reference evaluation, such as semi-simulated contamination.

\subsection{Adaptive Stability Under Non-stationarity}

Cutoff sensitivity analysis highlights operational implications of adaptive subspace tracking. The steep operating slope observed in HMO-ASR indicates high susceptibility to parameter mis-specification, limiting robustness under variable recording conditions. Original ASR, while stable, lacks responsiveness to evolving covariance structure.

EMA-ASR exhibits smoother operating transitions, indicating improved tolerance to parameter variation and gradual assimilation of non-stationary drift. This behavior reflects the stabilizing influence of dual-timescale adaptation, wherein rapid covariance evidence is tempered by slow recursive integration.

\section{Conclusion}

This work introduced an adaptive Artifact Subspace Reconstruction framework based on exponential moving average (EMA) subspace tracking. Dual-timescale adaptation enables continuous operation under intra-subject non-stationarity through a single adaptation-rate parameter. EMA-ASR achieved substantially stronger artifact suppression than static and memory-limited ASR variants, at the cost of increased reconstruction and spectral deviation, whereas HMO-ASR performed comparably to Original ASR at the chosen operating point. These results underscore the temporal fragility of calibration-defined subspaces and support lightweight adaptive tracking as a deployment-oriented approach to real-world EEG artifact removal. Future work will evaluate EMA-ASR against clean-reference ground truth and at signal degradation-matched operating points to disentangle artifact removal from attenuation of neural activity.

\section*{Code Availability}

MATLAB implementations of EMA-ASR, HMO-ASR, and the baseline ASR pipeline, together with the scripts used to generate all tables and figures, are publicly available at \url{https://github.com/NeuralLabIITGuwahati/EMA-ASR}. Parameter settings for all reported results are documented in the repository configuration.

\section*{Acknowledgment}

VKN is funded by the MoE doctoral scholarship from the Government of
India. CNG's time was funded by DST, Govt.\ of India (Project Code:
DST/INT/SWD/VR/P-14/2019).

\section*{Use of Generative AI}
The authors used large language models (LLMs) for language editing of the manuscript. The scientific content, results, and interpretations are the authors' own. The authors reviewed and verified all AI-assisted output and take full responsibility for the content of this article.

\bibliographystyle{IEEEtran}
\bibliography{refs}

\end{document}